\documentclass[10pt]{article}

\usepackage{graphics}
\usepackage{amsmath}
\usepackage{rotating}
\usepackage{cite}

\newlength{\figurewidth}
\renewcommand{\sectionmark}[1]{\markboth{\textsc{M. T. Gastner and
M. E. J. Newman}}{\textsc{Diffusion-based method for producing density
equalizing maps}}}
\renewcommand{\subsectionmark}[1]{\markboth{\textsc{M. T. Gastner and
M. E. J. Newman}}{\textsc{Diffusion-based method for producing density
equalizing maps}}}

\newcommand{\captionfonts}{\small}
\makeatletter
\long\def\@makecaption#1#2{%
  \vskip\abovecaptionskip
  \sbox\@tempboxa{{\captionfonts #1: #2}}%
  \ifdim \wd\@tempboxa >\hsize
    {\captionfonts #1: #2\par}
  \else
    \hbox to\hsize{\hfil\box\@tempboxa\hfil}%
  \fi
  \vskip\belowcaptionskip}
\makeatother

\renewcommand{\d}{{\rm d}}

\newcommand{\eref}[1]{(\ref{#1})}
\newcommand{\etal}{{\it{}et~al.}}
\newcommand{\del}{\nabla}
\newcommand{\vr}{\mathbf{r}}
\newcommand{\vk}{\mathbf{k}}
\newcommand{\vv}{\mathbf{v}}
\newcommand{\vJ}{\mathbf{J}}
\newcommand{\vT}{\mathbf{T}}

\begin{document}

\title{Diffusion-based method for producing\\
density equalizing maps}
\author{Michael T. Gastner and M. E. J. Newman\\
\\
\textit{\small Center for the Study of Complex Systems and Department of
Physics,}\\
\textit{\small University of Michigan, Ann Arbor, MI 48109}}
\date{}
\maketitle

\begin{abstract}
Map makers have long searched for a way to construct cartograms---maps in
which the sizes of geographic regions such as countries or provinces appear
in proportion to their population or some other analogous property.  Such
maps are invaluable for the representation of census results, election
returns, disease incidence, and many other kinds of human data.
Unfortunately, in order to scale regions and still have them fit together,
one is normally forced to distort the regions' shapes, potentially
resulting in maps that are difficult to read.  Many methods for making
cartograms have been proposed, some of them extremely complex, but all
suffer either from this lack of readability or from other pathologies, like
overlapping regions or strong dependence on the choice of coordinate axes.
Here we present a new technique based on ideas borrowed from elementary
physics that suffers none of these drawbacks.  Our method is conceptually
simple and produces useful, elegant, and easily readable maps.  We
illustrate the method with applications to the results of the 2000 US
presidential election, lung cancer cases in the State of New York, and the
geographical distribution of stories appearing in the news.
\end{abstract}

\section{Introduction}
Suppose we wish to represent on a map some data concerning, to take the
most common example, the human population.  For instance, we might wish to
show votes in an election, incidence of a disease, number of cars,
televisions, or phones in use, numbers of people falling in one group or
another of the population, by age or income, or any of very many other
variables of statistical, medical, or demographic interest.  The typical
course under such circumstances would be to choose one of the standard
projections for the area of interest and plot the data on it using some
color code or similar representation.  Such maps however can be highly
misleading.  A plot of disease incidence, for example, will inevitably show
high incidence in cities and low incidence in rural areas, solely because
there are more people living in cities.

The obvious cure for this problem is to plot a fractional measure rather
than raw incidence data; we plot some measure of the number of cases per
capita, binned in segments small enough to give good spatial resolution but
large enough to give reliable sampling.  This however has its own problems,
since it discards all information about where most of the cases are
occurring.  One case in a thousand means something entirely different in
Sydney from what it means in Siberia.

What we would like is some representation of the data that factors out
variations in the population density but, at the same time, shows how many
cases are occurring in each region.  It appears at first that these two
goals are irreconcilable, but this is not the case.  On a normal
area-preserving or approximately area-preserving projection, such as a
Mercator or Robinson projection, they are indeed irreconcilable.  However,
if we can construct a projection in which areas on the map are proportional
not to areas on the ground but instead to human population (or whatever the
corresponding variable might be for the problem in hand), then we can have
our cake and eat it.  Disease cases or other similar data plotted on such a
projection will have the same density in areas with equal per capita
incidence regardless of the population, since both the raw incidence rate
and the area will scale with the population.  However, each case or group
of cases can still be represented individually, so it will be clear to the
eye where most of the cases occur.  Projections of this kind are known as
value-by-area maps, density equalizing maps, or cartograms.

The construction of cartograms turns out to be a challenging undertaking.
A variety of methods have been put forward, but none is entirely
satisfactory.  In particular, they often produce highly distorted maps that
are difficult to read or projections that are badly behaved under some
circumstances, with overlapping regions or strong dependence on coordinate
axes.  In many cases the methods proposed are also computationally
demanding, sometimes taking hours to produce a single map.  In this paper
we propose a new method which is, we believe, intuitive, while also
producing elegant, well-behaved, and useful cartograms, whose calculation
makes relatively low demands on our computational resources.

\section{Previous methods for constructing cartograms}
\label{others}
Before introducing our own method for constructing cartograms, we outline
in this section some of the previous methods that have been proposed.  We
will in most of the paper be considering cartograms based on population
density, although the developments are easily generalized to other
quantities.  (We give one different example in Sec.~\ref{examples}.)  The
construction of a (flat two-dimensional) cartogram involves finding a
transformation $\vr\to\vT(\vr)$ of a plane to another plane such that the
Jacobian $\partial(T_x,T_y)/\partial(x,y)$ of the transformation is
proportional to some specified (population) density $\rho(\vr)$ thus:
\begin{equation}
{\partial(T_x,T_y)\over\partial(x,y)} \equiv
\frac{\partial T_x}{\partial x}\,\frac{\partial T_y}{\partial y} -
\frac{\partial T_x}{\partial y}\,\frac{\partial T_y}{\partial x}
  = {\rho(\vr)\over\bar{\rho}},
\label{jacobian}
\end{equation}
where $\bar{\rho}$ is the mean population density averaged over the
area~$A$ to be mapped.  (This choice of normalization for the Jacobian
ensures that the total area before and after the transformation is the
same.)

Eq.~\eref{jacobian} does not determine the cartogram projection uniquely.
To do that we need one more constraint: two constraints are needed to fix
the projection for a two-dimensional cartogram.  Different choices of the
second constraint give different projections, and no single choice appears
to be the obvious candidate, which is why many methods of making cartograms
have been suggested.  One idea is to demand conformal invariance under the
cartogram transformation, i.e.,~to demand that angles be preserved locally.
This is equivalent to demanding that the Cauchy--Riemann equations be
satisfied, but this imposes two, not one, additional constraints and hence
it is normally not possible to construct a conformally invariant cartogram.

In an attempt at least to minimize the distortion of angles,
Tobler~\cite{T63,T73} took the first steps in the automated computer
generation of cartograms in the late 1960s.  He proposed a method in which
the initial map is divided into small rectangular or hexagonal cells, each
of which is then independently dilated or shrunk to a size proportional to
its population content.  Since each cell is scaled separately, the corners
of adjacent cells don't match afterwards.  To re-establish a match,
Tobler's method takes a vector average over the positions of corresponding
corners and draws a new map with the resulting distorted cells.  The
process is iterated until a fixed point of the transformation is reached.
Although the principle is simple and intuitive it runs into practical
problems.  First, convergence tends to be rather slow because a node a few
cells away from a population center will feel the effect of that center
only after several iterations.  Second, under some circumstances the
transformation can produce overlapping or ``folded'' regions of the map,
thereby ruining the topology.  This problem can be corrected by introducing
additional constraints but the result is a more complex algorithm with even
slower run-times.

To increase the speed of the calculations, Dougenik~\etal~\cite{DCN85}
introduced an algorithm where the borders of a cell move in response not
only to local space requirements but also to ``forces'' exerted by other
cells.  Cells create force fields that diminish with distance from the cell
and that are larger for cells that contain larger populations.  These
forces ``push'' other cells away from areas of high population in a manner
reminiscent of the behavior of charged objects in electrostatics (although
the authors do not use this metaphor).  Again the positions are relaxed
iteratively to achieve the final cartogram, and convergence is
substantially faster than Tobler's algorithm, although topological errors
still cannot be ruled out.

Gusein-Zade and Tikunov~\cite{GT93} suggested a further twist that does
away with the cells altogether and uses a continuous ``displacement field''
that measures the displacement of each point in the map.  Areas of high
population exert a repulsive force on this displacement field and the
authors are able to derive a differential equation for the field, which
they integrate numerically.  The method is slow and somewhat arcane but
produces some of the most attractive cartograms among the existing
algorithms.

A more intuitive method, which makes use of cellular automata, has recently
been proposed by Dorling~\cite{D96}.  In this method the original map is
drawn on a fine grid.  On each iteration of the algorithm, cells lying on
or close to the boundaries of regions are identified and if a neighboring
region needs extra area those cells are reassigned to the neighbor.  The
procedure is iterated and the regions with greatest population grow slowly
larger until an equilibrium is reached and no further changes are needed.
The procedure is elegant and simple, but in practice it can distort shapes
quite badly.  One can add additional constraints on the shapes to make the
maps more readable, but then the method quickly loses its main advantage,
namely its simplicity.

Researchers have also experimented with several other methods.
Kocmoud~\cite{K97}, for example, uses a mass-and-spring model acting on a
map expressed as points and lines, with constraints applied to maintain
certain topographic features such as angles or lengths.  Due to its
complexity, however, this algorithm is prohibitively slow.  The method of
Keim~\etal~\cite{K03}, by contrast, is very fast, but achieves its speed
primarily by working with polygonal maps that have been heavily simplified
before beginning the computations, which unfortunately dispenses with many
useful cartographic details.  Finally, if one is willing to live with a
non-contiguous cartogram---one in which regions adjacent in real life are
not adjacent on the cartogram---then several quite simple methods give good
results, such as Dorling's circular cartograms~\cite{D96}.  Other reviews
and discussions of cartogram methods can be found in
Refs.~\cite{SMSWBS84,CS89,T90,EW97}.

\section{The diffusion cartogram}
\label{diffcart}
In this paper we propose a new type of cartogram, which might be called a
``diffusion cartogram,'' for reasons we now describe.  It is a trivial
observation that on a true population cartogram the population is
necessarily uniform: once the areas of regions have been scaled to be
proportional to their population then, by definition, population
\emph{density} is the same everywhere---hence the name ``density-equalizing
map.''  Thus, one way to create a cartogram given a particular population
density, is to allow population somehow to ``flow away'' from high-density
areas into low-density ones, until the density is equalized everywhere.
There is an obvious candidate process that achieves this, the linear
diffusion process of elementary physics~\cite{HR93}, and this is the basis
of our method.  We describe the population by a density
function~$\rho(\vr)$, where $\vr$ represents geographic position, and then
allow this density to diffuse.  In the limit of long time $t\to\infty$, the
population density becomes uniform and so comes to rest, and its total
displacement from start to finish determines the projection of the map
necessary to produce a perfectly density-equalizing cartogram.

In the standard treatment of diffusion, the current density is given by
\begin{equation}
\vJ = \vv(\vr,t)\,\rho(\vr,t),
\label{defsv}
\end{equation}
where $\vv(\vr,t)$ and $\rho(\vr,t)$ are the velocity and density
respectively at position~$\vr$ and time~$t$.  Diffusion follows the
gradient of the density field thus:
\begin{equation}
\vJ = -\del\rho,
\label{defsj}
\end{equation}
meaning that the flow is always directed from regions of high density to
regions of low and that the flow will be faster when the gradient is
steeper.  There is conventionally a diffusion constant in Eq.~\eref{defsj},
but we can set this constant to~1 without affecting our results.

The diffusing population is also conserved locally so that
\begin{equation}
\del\cdot\vJ + \frac{\partial\rho}{\partial t} = 0.
\label{conserve}
\end{equation}
Combining Eqs.~\eref{defsv}, \eref{defsj}, and~\eref{conserve} we then
arrive at the familiar diffusion equation:
\begin{equation}
\del^2\rho - \frac{\partial\rho}{\partial t} = 0
\label{diffusion}
\end{equation}
and the expression for the velocity field in terms of the population
density:
\begin{equation}
\vv(\vr,t) = -\frac{\del\rho}{\rho}.
\label{gradlaw}
\end{equation}

The calculation of the cartogram involves solving Eq.~\eref{diffusion} for
$\rho(\vr,t)$ starting from the initial condition in which $\rho$ is equal
to the given population density of the region of interest and then
calculating the corresponding velocity field from Eq.~\eref{gradlaw}.  The
cumulative displacement $\vr(t)$ of any point on the map at time~$t$ can be
calculated by integrating the velocity field thus:
\begin{equation}
\vr(t) = \vr(0) + \int_0^t \vv(\vr,t') \>\d t'.
\label{displacement}
\end{equation}
In the limit $t\to\infty$ the set of such displacements for all points on
the original map defines the cartogram.

Most of the time we are not interested in mapping the entire globe, but
only some part of it, which means that the area of interest will have
boundaries beyond which we don't know or don't care about the population
density.  These boundaries might be country borders or other political
boundaries or coast lines.  It would be inappropriate to represent regions
outside the boundaries as having zero population, even if they are, like
the ocean for instance, unpopulated, since this would cause arbitrary
expansion of the cartogram as the population diffused into its uninhabited
surroundings.  (This is true of essentially all methods for constructing
cartograms.)  Instead, therefore, we apply a ``neutral buoyancy''
condition, floating the area of interest in a ``sea'' of uniform population
density equal to the mean density of the map as a whole.  This keeps the
total area under consideration constant during the diffusion process.

The whole simulation, including the sea, is then enclosed in a box.  For
simplicity in this paper we will consider only rectangular boxes, as most
others have also done.  Provided the dimensions $L_x$ and $L_y$ of the box
are substantially larger than the area to be mapped, the dimensions
themselves do not matter.  In the limit $L_x,L_y\to\infty$ the cartogram
will be a unique deterministic mapping, independent of the coordinate
system used, with no overlapping regions.  In practice, we find that quite
moderate system sizes are adequate---dimensions three to four times the
linear extent of the area to be mapped appear to give good results.

We also need to choose boundary conditions on the walls of the box.  These
too have no great effect on the results, provided the size of the box is
reasonably generous, and we have found a good choice to be the Neumann
boundary conditions in which there is no flow of population through the
walls of the box.

The considerations above completely specify our method and are intuitive
and straightforward.  The actual implementation of the method, if one wants
a calculation that runs quickly, involves a little more work.  We solve the
diffusion equation in Fourier space, where it is diagonal, and
back-transform before integrating over the velocity field.  With the
Neumann boundary conditions, the appropriate Fourier basis is the cosine
basis, in which the solution to the diffusion equation has the form
\begin{equation}
\rho(\vr,t) = 
\frac{4}{L_x L_y}\sum_{\vk}
  \tilde{\rho}(\vk) \cos(k_x x) \cos(k_y y) \exp(-k^2 t),
\label{rho}
\end{equation}
where the sum is over all wavevectors $\vk=(k_x,k_y)=2\pi(m/L_x,n/L_y)$
with $m$, $n$ non-negative integers, and $\tilde\rho(\vk)$ is the discrete
cosine transform of $\rho(\vr,t=0)$:
\begin{equation}
\tilde{\rho}(\vk) = 
  \begin{cases}
    \frac{1}{4}\int_0^{L_x}\int_0^{L_y} \rho(\vr,0)\>\d x\,\d y & 
      \text{if $k_x=k_y=0$,} \\
    \frac{1}{2}\int_0^{L_x}\int_0^{L_y} \rho(\vr,0)\cos(k_y y)\>\d x\,\d y & 
      \text{if $k_x=0$ and $k_y \neq 0$,} \\
    \frac{1}{2}\int_0^{L_x}\int_0^{L_y} \rho(\vr,0)\cos(k_x x)\>\d x\,\d y & 
      \text{if $k_x \neq 0$ and $k_y=0$,} \\
    \phantom{\frac{1}{2}}\int_0^{L_x}\int_0^{L_y}
      \rho(\vr,0)\cos(k_x x)\cos(k_y y)\>\d x\,\d y \quad & \text{otherwise.}
  \end{cases}
\label{rhoft}
\end{equation}
The velocity field $\vv$ is then easily calculated from Eqs.~\eref{gradlaw}
and~\eref{rho} and has components
\begin{subequations}
\label{velocity}
\begin{eqnarray}
v_x(\vr,t) &=& 
\frac{\sum_\vk k_x\tilde\rho(\vk)\sin(k_x x)\cos(k_y y)\exp(-k^2 t)}%
{\sum_\vk \tilde\rho(\vk)\cos(k_x x)\cos(k_y y)\exp(-k^2 t)}, \\
v_y(\vr,t) &=& 
\frac{\sum_\vk k_y\tilde\rho(\vk)\cos(k_x x)\sin(k_y y)\exp(-k^2 t)}%
{\sum_\vk \tilde\rho(\vk)\cos(k_x x)\cos(k_y y)\exp(-k^2 t)}.
\end{eqnarray}
\end{subequations}
Equations~\eref{rhoft} and~\eref{velocity} can be evaluated rapidly using
the fast Fourier transform (FFT) and its back-transform respectively, both
of which in this case run in time of order $L_x L_y \log (L_x L_y)$.  We
then use the resulting velocity field to integrate Eq.~\eref{displacement},
which is a nonlinear Volterra equation of the second kind and can be solved
numerically by standard methods~\cite{PTVF92}.  In practice it is the
Fourier transform that is the time-consuming step of the calculation and,
as we will see, with the aid of the FFT this step can be performed fast
enough that the whole calculation runs to completion in a matter of seconds
or at most minutes, even for large and detailed maps.

It is straightforward to see that our diffusion cartogram satisfies the
fundamental definition, Eq.~\eref{jacobian}, of a cartogram.  In the limit
$t\to\infty$, Eq.~\eref{rho} is dominated by the $\vk=0$ term and gives
\begin{equation}
\rho(\vr,\infty) = {4\tilde{\rho}(0)\over L_x L_y}
  = {1\over L_xL_y} \int_0^{L_x}\int_0^{L_y} \rho(\vr,0)\>\d x\,\d y
  = \bar{\rho},
\end{equation}
where $\bar{\rho}$ is again the mean population density over the area
mapped.  Furthermore, by definition, the total population within any moving
element of area does not change during the diffusion process, and hence,
denoting by $\vT(\vr)$ the final position of a point that starts at
position~$\vr$, we have $\rho(\vr)\>\d x\,\d y = \bar{\rho}\>\d T_x\,\d
T_y$, and, rearranging, the Jacobian is given by
$\partial(T_x,T_y)/\partial(x,y) = \rho(\vr)/\bar{\rho}$, in agreement with
Eq.~\eref{jacobian}.

Below we present a number of applications of our cartograms to the
visualization of various types of data.  Before we do that however, it is
worth giving a brief comparison of our method with the others discussed in
Sec.~\ref{others}.  The main advantages of our method over the original
method of Tobler~\cite{T63,T73}, are first speed and second independence of
a particular choice of cells.  Tobler's method is obliged to choose
directions for the dilation of the cells and this means that the resulting
cartogram depends on the coordinate axes used.  As we will see, the
practical application of our algorithm, like most numerical calculations,
relies on a sampling of the population density on a lattice of some kind,
but the diffusion process is, in the limit of small lattice spacing,
independent of the topology of this lattice and therefore does not show
lattice-dependent effects like the method of Tobler.  The method of
Gusein-Zade and Tikunov~\cite{GT93} is also free of these effects, but pays
for it in run time and complexity.

Conceptually our algorithm is in some respects similar to the cellular
automaton method of Dorling~\cite{D96}.  Our description of the diffusion
method has been entirely in terms of macroscopic variables and equations,
but one could equally look at the method as a microscopic diffusion process
in which each individual member of the population performs a Gaussian
random walk about the surface of the map.  Over time the population will
diffuse until it is uniform everywhere within the box enclosing the map,
except for statistical fluctuations.  The projection of the cartogram is
then derived by moving all boundaries on the map in such a way that the net
flow passing through them is zero at all times during the diffusion
process.  This resembles Dorling's method in the sense that different
regions trade their area until a fair distribution is reached.

Our method however has an advantage over Dorling's in being based on a
global, lattice-independent diffusion process.  The exchange of area
between regions in Dorling's method occurs only between nearest-neighbor
squares along the principle axes of a square lattice and this introduces a
strong signature of the lattice topology into the final cartogram.

\section{Population density function}
\label{popdensity}
The description of our method tells, in a sense, only half the story of how
to create a cartogram.  Before applying this or indeed any method, we need
to choose the starting density $\rho(\vr)$ for the map.  We can, by
defining $\rho(\vr)$ in different ways, control the properties of the
resulting cartogram, including crucially the balance between accurate
density equalization and readability.

Population density is not strictly a continuous function, since people are
themselves discrete and not continuous.  To make a continuous function the
population must be binned or coarse-grained at some level.  All methods of
constructing cartograms require one to do this and there is no single
accepted standard approach.  Part of the art of making a good cartogram
lies in shrewd decisions about the definition of the population density.

If we choose a very fine level of coarse-graining for the population
density, then the high populations in centers such as cities will require
substantial local distortions of the map in order to equalize the density.
A coarser population density will cause less distortion, resulting in a map
with features that are easier to recognize, but will give a less accurate
impression of the true population distribution.  The most common choice
made by others has been to coarse-grain the population at the level of the
(usually political) regions of interest.  For example, if one were
interested in the United States, one might take the population of each
state and distribute it uniformly over the area occupied by that state.
This method can be used also with our cartogram algorithm and we give one
example below.  But we are not obliged to use it, and in some cases it may
be undesirable, since binning at the level of states erases any details of
population distribution below the state level.  In our work we have mostly
adopted instead a spatially uniform Gaussian blur as our coarse-graining
function.  By varying the width~$\sigma$ of the blurring function we can
tune our cartogram between accuracy and readability.

Ultimately the choice of population density function is up to the user of
the method, who must decide what particular features are most desirable in
his or her application.  One advantage of our diffusion-based algorithm is
that it is entirely agnostic about this choice; the process of computing
the cartogram is completely decoupled from the calculation of the
population density and hence is not slanted in favor of one choice or
another.

\section{Applications}
\label{examples}
We give three examples of the use of our cartograms, focusing on the United
States and using population data from the 2000 US Census.

\begin{figure}
\begin{center}
\resizebox{\textwidth}{!}{\includegraphics{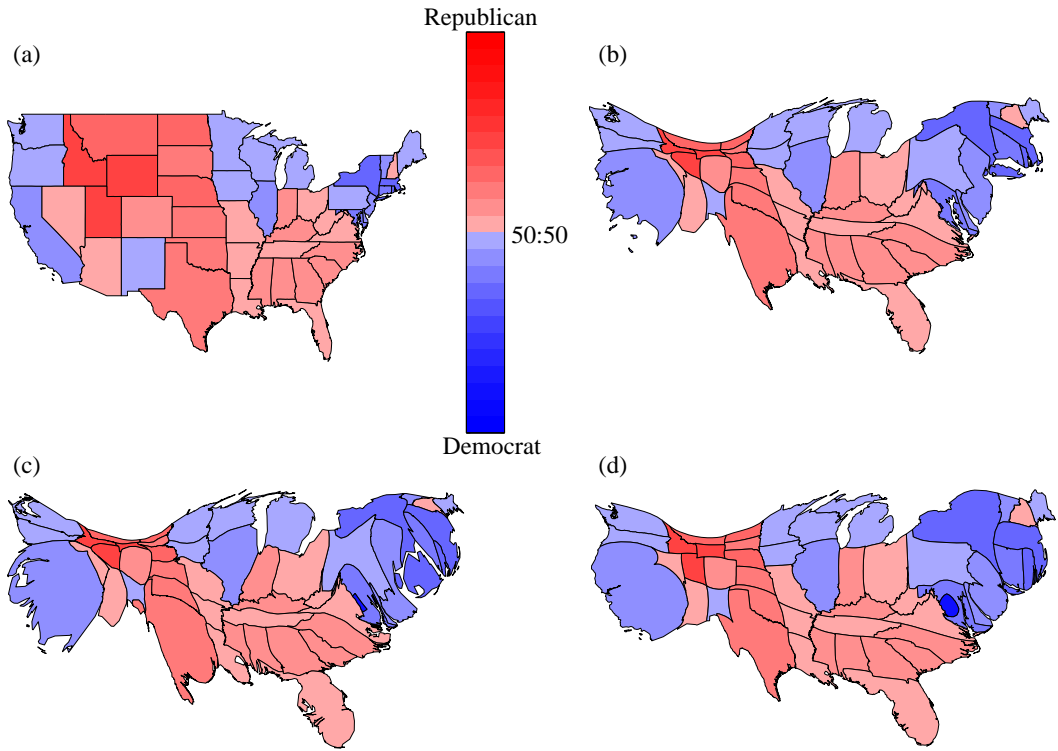}}
\end{center}
\caption{The results of the 2000 US presidential election.  (a)~A normal
cylindrical projection.  (b)~A cartogram with moderate coarse-graining
generated using a Gaussian blur of standard deviation $\sigma =
0.75^\circ$, corresponding to about 80km.  (c)~A cartogram constructed
using a narrower Gaussian ($\sigma = 0.1^\circ$), resulting in greater
distortion of state boundaries especially in the northeast---most
noticeably for Pennsylvania and Massachusetts.  (d)~A cartogram based on
states' representation in the electoral college.  The density of electors
was calculated by spreading each state's electors evenly across the state.}
\label{election}
\end{figure}

First, we examine the results of the US presidential election of 2000.
Figure~\ref{election}a shows the popular vote by state in this
election---the simple fraction of voters voting each way---with state
majorities for the two major candidates represented as shades of red
(George Bush, Republican) and blue (Al Gore, Democrat).  The briefest
appraisal by eye immediately reveals that the Republicans dominate much
more than a half of the country.  This, however, is a misleading conclusion
because much of the Republicans' dominance comes from their success in the
large but relatively unpopulated states in the center of the map, while the
Democrats carry the more populous areas in the northeast and on the west
coast.  Clearly then, this is a poor visual representation of the results,
in the sense that it is hard to tell which party got more votes by looking
at the map.

A much better representation is given by Figure~\ref{election}b, in which
the same color code is used on a population cartogram constructed using the
methods described in this paper.  This cartogram is based on a density
function with a moderately broad Gaussian blur, as described above,
yielding a map whose features are distorted relatively little on short
scales; the individual states are still easily recognizable while being
scaled close to the size appropriate to their populations.  To a good
approximation the amounts of red and blue in the figure now correspond to
the true balance of the popular vote, and, as is clear to the eye, this
vote was very close between the two parties, in fact being won not by the
Republican candidate but by the Democrat.  In Fig.~\ref{election}c, we show
a further cartogram constructed using less coarse-graining on the
population density, resulting in a map that more perfectly represents
states' populations, but which also has more distortion on short
length-scales, making some regions hard to recognize.  For example, the
small but densely populated Long Island now expands (quite correctly) to a
size greater than the entire state of Wyoming.  The user concerned both
with readability and accurate portrayal of the data would probably choose a
map similar to Fig.~\ref{election}b in this case.

Ultimately, the presidency is decided not by the popular vote, but by the
electoral college.  Under the US system, each state contributes a certain
number of electors to the electoral college, who vote according to the
majority in their state.  The candidate receiving a majority of the votes
in the electoral college wins the election.  The appropriate visualization
for a vote of this kind is one in which the sizes of the states are scaled
in proportion to their numbers of electors.  This then is an example in
which a coarse-graining according to political boundaries (state boundaries
in this case) makes good sense.  We show a cartogram calculated in this way
in Fig.~\ref{election}d.  The allocation of electors to states roughly
follows population levels, but contains a deliberate bias in favor of less
populous states, and as a result some of these states appear larger in
Fig.~\ref{election}d than in~\ref{election}c---Wyoming, Montana, and the
Dakotas are good examples.  Since most of these are majority-Republican
states, we can now understand how the Republican candidate came to win the
election in spite of losing the popular vote.

\begin{figure}
\begin{center}
\resizebox{\textwidth}{!}{\includegraphics{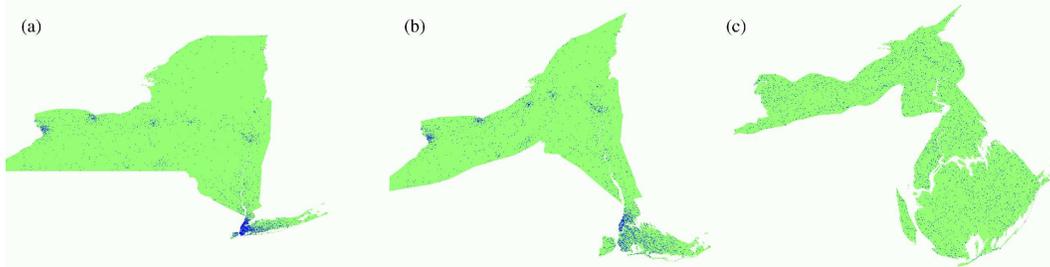}}
\end{center}
\caption{Visualization of lung cancer cases among males in the state of New 
York 1993--1997.  Each dot represents ten cases, randomly placed within the
zip-code area of occurrence.  (a)~The original map.  (b)~A cartogram using
a coarse-grained population density with $\sigma = 0.3^\circ$.  (c)~A
cartogram using a much finer-grained population density
($\sigma=0.04^\circ$).  (Data from the New York State Department of
Health.)}
\label{cancer}
\end{figure}

For our second example, we look at a case in which a very fine level of
coarse-graining is needed to understand the data fully.  We study the
distribution of cases of lung cancer among the male population in the state
of New York.  (A similar study using a different technique and for a
smaller area was carried out by Merrill~\cite{M01}.)  In Fig.~\ref{cancer}a
we show a scatter plot of lung cancer cases between 1993 and 1997, as
recorded by the New York State Department of Health.  This map is of
precisely the kind discussed in the introduction: it shows clearly how many
cases there are, and that they are geographically concentrated in the areas
that have high populations.  However, it is impossible to tell whether
there is a statistically higher per capita incidence of lung cancer in one
area or another, because any such variation is masked by the highly
non-uniform population density.

In Fig.~\ref{cancer}b, we show the same data on a population cartogram with
moderate coarse-graining of the initial population density.  Although the
map is visibly distorted, little difference is visible in the distribution
of cancer cases.  In Fig.~\ref{cancer}c, on the other hand, we use a very
fine-grained population density, creating a cartogram with better
population equalization and significantly greater distortion.  Now the
virtue of this representation becomes strikingly clear.  As the figure
shows, when we use a projection that truly equalizes the population density
over the map, there is no longer any significant variation in the
distribution of cases over the state---the dots have about the same density
everywhere.  The shape of the map in (c) does not much resemble the shape
of the original any more, but this is the price we pay for equalizing the
population almost perfectly.

Our method of generating cartograms is fast, an important consideration for
interactive use.  As discussed in Sec.~\ref{diffcart}, the bulk of the work
involved in creating the maps is in the Fourier transforms, which can be
computed rapidly using FFTs.  Fig.~\ref{cancer}c for example was produced
in less than 100 seconds on a standard desktop computer, including the time
to read in the census data, to perform the Gaussian convolution, to solve
the diffusion equation, and to plot the figure.  Previous techniques are
either significantly slower (Kocmoud~\cite{K97} reports 16 hours for a US
state cartogram using his constraint-based approach) or are obliged to use
simplified maps to reduce the computational load~\cite{K03}.

The example cartograms given so far have all been based, more or less, on
human population density, which is certainly the most common type of
cartogram.  Other types however are also possible and for our third example
we study one such.  Anyone who reads or watches the news in the United
States (and similar observations probably apply in other countries as well)
will have noticed that the geographical distribution of news stories is
skewed.  Even allowing for population, a few cities, notably New York and
Washington, DC, get a surprisingly large fraction of the attention while
other places get little.  Apparently some locations loom larger in our
mental map of the nation than others, at least as presented by the major
media.  We can turn this qualitative idea of a mental map into a real map
using our cartogram method.

\begin{figure}
\begin{center}
\resizebox{!}{11cm}{\includegraphics{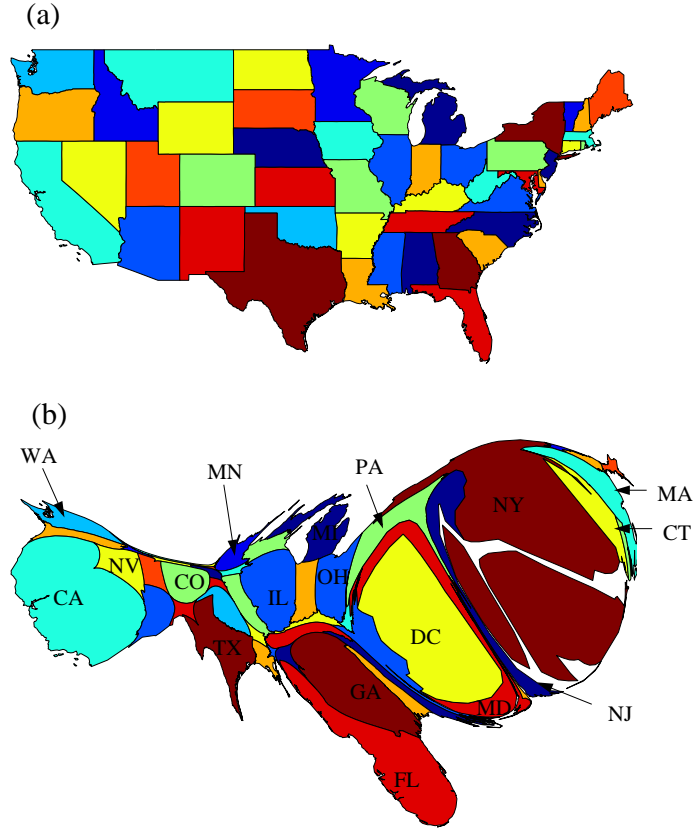}}
\end{center}
\caption{The distribution of news stories by state in the United States.
(a)~Conventional map of the states.  (b)~``Mindshare map'' in which the
sizes of states are proportional to the frequency of their appearance in
news stories.  States have the same colors in (a) and~(b).}
\label{apws}
\end{figure}

We have taken about $72\,000$ newswire stories from November 1994 to April
1998, and extracted from each the ``dateline,'' a line at the head of the
story that gives the date and the location that is the main focus of the
story.  Binning these locations by state, we then produce a ``mindshare
map'' in which the sizes of the US states represent the fraction of stories
concerning that state over the time interval in question.  The result is
shown in Fig.~\ref{apws}.

The stories are highly unevenly distributed.  New York City alone
contributes $20\,000$ stories to the corpus---largely because of the
preponderance of wire stories about the financial markets---and Washington,
DC another $10\,000$, largely political stories.  We chose to bin by state
to avoid large distortions around the cities that are the focus of most
news stories.  We made one exception however: since New York City had far
more hits than any other location including the rest of the state of New
York (which had around 1000), we decided to split New York State into two
regions, one for the greater New York City area and another for the rest of
the state.

The cartogram is a dramatic depiction of the distribution of US news
stories.  The map is highly distorted because the patterns of reporting
show such extreme variation.  Washington, DC, for instance, which is
normally invisible on a map of this scale, becomes the second largest
``state'' in the union.  (The District of Columbia is not, technically, a
state.)  Members of the public, including the US public, frequently
overestimate the size of the northeastern part of the United States by
comparison with the middle and western states, and this map may give us a
clue as to why.  Perhaps people's mental image of the United States is not
really an inaccurate one; it is simply based on things---such as
``mindshare''---that people find more relevant than geographical area.

\section{Conclusions}
\label{concs}
In this paper we have presented a new general method for constructing
density-equalizing projections or cartograms, which provide an invaluable
tool for the presentation of geographic data.  Our method is simpler than
many earlier methods, allowing for rapid calculations, while generating
accurate and readable maps.  The method allows its users to choose their
own balance between good density equalization and low distortion of map
regions, making it flexible enough for a wide variety of applications.  We
have given comparisons between our method and previous ones, and presented
a number of examples of the use of our cartograms in the representation of
human data, including election results and incidence data for cancer.

Implementation of our method in GIS software packages should be
straightforward, and we hope that in this or other forms it will prove a
valuable tool for researchers in a wide variety of disciplines.

\section*{Acknowledgments}
The authors would like to thank the staff of the University of Michigan's
Numeric and Spatial Data Services for their help with the geographic data,
and Dragomir Radev for useful discussions about the geographic distribution
of news messages.  This work was funded in part by the National Science
Foundation under grant number DMS--0234188 and by the James S. McDonnell
Foundation.

\end{document}